\documentclass[twocolumn,showpacs,preprintnumbers,amsmath,amssymb]{revtex4}

\usepackage{graphicx}

\begin{document}
\title{Theory for Photon-Assisted Macroscopic Quantum Tunneling 
in a Stack of Intrinsic Josephson Junctions}
\author{Masahiko Machida$^{a,c}$ and Tomio Koyama$^{b,c}$}
 \affiliation{
$^{a}$CCSE, Japan Atomic Energy Agency, 6-9-3 Higashi-Ueno, Taito-ku
Tokyo 110-0015, Japan  \\ 
$^{b}$IMR, Tohoku University, 2-1-1 Katahira Aoba-ku, Sendai 980-8577, Japan \\
$^{c}$CREST(JST), 4-1-8 Honcho, Kawaguchi, Saitama 332-0012, Japan}

\begin{abstract}
We propose a theory for photon-assisted macroscopic quantum tunneling 
(MQT) in a stack of capacitively-coupled intrinsic Josephson junctions 
in which the longitudinal Josephson plasma, i.e., longitudinal collective 
phase oscillation modes, is excited. The scheme of energy-level quantization 
in the collective oscillatory states is clarified in the $N$-junction system. 
When the MQT occurs from the single-plasmon states excited by microwave 
irradiation in the multi-photon process to the uniform voltage state, our 
theory predicts that the escape rate is proportional to $N^2$. This result is 
consistent with the recent observation in Bi-2212 intrinsic Josephson 
junctions.
 
\end{abstract}

\pacs{74.50.+r, 74.81.Fa, 74.72.Hs, 74.25.Nf}
\maketitle

Since the discovery of macroscopic quantum tunneling (MQT) in YBCO grain 
boundary junctions \cite{Bauch} and Bi-2212 intrinsic Josephson junctions 
(IJJs) \cite{Inomata}, a renewed interest has been aroused on the MQT in 
the Josephson effect from viewpoints of both basic physics and 
quantum-devise applications. In Bi-2212 IJJs the observed crossover 
temperature at which MQT appears is about 1K\cite{Inomata}, which is 1-order 
higher than that in conventional single-junction systems \cite{Voss, Devoret}, 
and, furthermore, 
the dissipation effect that prevents a quantum tunneling is very weak in spite 
that the order parameter of the high-$T_c$ cuprates has the d-wave symmetry with nodes. 
This is because the Josephson plasma frequency in IJJs is far higher
compared with artificially-made S/I/S Josephson junctions \cite{Kawabata}
and the transfer-integral along the $c$-direction vanishes for the nodal
quasi-particles. These remarkable features in IJJs give great promise for quantum-devise 
applications. In addition, MQT in IJJs is expected to provide a new physics 
originating from the atomic-scale multi-junction structure. 

Recently, Jin et al. observed MQT in the switching events to the uniform 
voltage state in Bi-2212 IJJs where all the junctions are switched into 
the voltage state\cite{Jin}. The switching rate observed in the multi-photon 
process is greatly enhanced, depending on the number of stacked junctions, 
$N$. The observed enhancement in the collective switching is proportional to 
$N^2$, which suggests that collective motion of the phase differences 
in IJJs is responsible for the quantum tunneling. In IJJs the longitudinal 
Josephson plasma is known to exist as the collective motion of the phase 
differences that propagates in the stacking direction of the junctions 
\cite{Koyama1}. In this paper we formulate a theory for MQT of the collective 
longitudinal plasma modes. The scheme of quantum energy levels for the 
collective motion of the phase differences in capacitively-coupled $N$ 
intrinsic junctions is clarified. Our theory predicts that the switching 
rate in the multi-photon process is proportional to $N^2$.


Consider a stack of intrinsic Josephson junctions having tiny in-plane area 
of $W \sim 1\mu m^2$. In the absence of an external magnetic field 
the phase of the order parameter $\varphi_\ell$, $\ell$ being the layer 
index, can be considered uniform along the in-plane direction. 
In this case the coupling between junctions in equilibrium state is brought 
about by the long-range Coulomb interaction between charges induced in the 
superconducting layers with an atomic-scale layer thickness as discussed 
in \cite{Koyama1}, that is, the coupling originates from the incomplete charge 
screening between junctions. The charge and phase dynamics in this case is 
well described by the Lagrangian \cite{Machida1},  
$$
\mathcal{L}=W\sum_{\ell=1}^\infty\Bigl\{\frac{s}{8\pi\mu^2}\bigl(A_\ell^0
+\frac{\hbar}{e^\ast}\dot\varphi_\ell\bigl)^2
+\frac{\epsilon d}{8\pi}E_{\ell,\ell-1}^2
$$
\begin{equation}
-E_J[1-\cos(\varphi_\ell-\varphi_{\ell-1})\bigl]\Bigl\},\label{(1)}
\end{equation}
where $A_\ell^0$ is the scalar potential at the $\ell$th superconducting 
layer, $E_{\ell,\ell-1}$ is the electric field inside the insulating layer 
with the dielectric constant $\epsilon$ between $\ell$th and $(\ell-1)$th 
superconducting layers, and $s$ and $d$ are, respectively, the thicknesses 
of the superconducting and insulating layers. The Josephson coupling energy 
in Eq.(1) is defined as $E_J=\hbar j_c/|e^\ast|$, $j_c$ being the Josephson 
critical current density. The first term including $A_\ell^0$ in Eq.(1) 
corresponds to the charging energy which leads to the finite charge 
compressibility \cite{Marel1}. To see this we note that the canonical 
momentum which is conjugate to $\varphi_\ell$ is given as  
\begin{equation}
p_\ell=\frac{\partial\mathcal{L}}{\partial\dot\varphi_\ell}
=\frac{\hbar}{e^\ast}\cdot\frac{s}{4\pi\mu^2}\bigl(A_\ell^0
+\frac{\hbar}{e^\ast}\dot\varphi_\ell\bigl)=\hbar sn_\ell. \label{(2)}
\end{equation}
where $n_\ell$ is understood to be the density fluctuation of the Cooper-pairs 
in the $\ell$th layer. Then, the first term in Eq.(1) increases the free 
energy by the amount of $\propto n_\ell^2$, which leads to the finite charge 
compressibility discussed by van der Marel and Tsvetkov \cite{Marel1}. The 
effect of the charge compressibility has been observed in optical properties 
of high-$T_c$ superconductors \cite{Dulic, Kakeshita}. 

The Lagrangian (1) yields the Hamiltonian of the form\cite{Koyama2}, 
\begin{equation}
\mathcal{H}=\frac{1}{2}\sum_{\ell m}C_{\ell m}^{-1}Q_\ell Q_m 
+E_J\sum_\ell\bigl[1-\cos(\varphi_\ell-\varphi_{\ell-1})\bigl],\label{(3)}
\end{equation}
where $Q_\ell$ is the total charge of the $\ell$th layer, i.e., $Q_\ell=
e^\ast Wsn_\ell$, and $C_{\ell m}^{-1}$ is the inverse capacitance matrix. 
Note that the condition $d^2\ll W$ is fulfilled even in the intrinsic 
Josephson junctions with a small in-plane area of $W\sim 1\mu$m$^2$, since 
$d\simeq 1.2$nm. From this fact one understands that the electric field 
generated in the junctions is well confined inside the stack of the 
junctions and its direction is perpendicular to the junctions, and as a result, the 1D Coulomb potential is realized almost
completely. Then, the inverse mutual capacitance diverges linearly as\cite{Marel1}
\begin{equation}
C_{\ell m}^{-1}\rightarrow \frac{4\pi d}{\varepsilon W}\ell, \ \ \ \ 
{\rm for} \ \ell\gg m. \label{(4)}
\end{equation}
This result indicates that the long-range nature of the capacitive coupling 
between junctions should be correctly incorporated, that is, any truncation 
of the inverse capacitance matrix in Eq.(3) cannot be accepted in the 
intrinsic Josephson junctions. As shown in \cite{Koyama2}, the Hamiltonian 
(3) can be transformed into more tractable form by the canonical 
transformation. 
We choose $(\theta_\ell,u_\ell)$ defined as 
\begin{equation}
 \theta_\ell=\varphi_\ell-\varphi_{\ell-1},  \ \ \ \ 
 u_\ell=\sum_{m=\ell}^\infty p_m,\label{(5)}
\end{equation}
as the canonical variables instead of $(\varphi_\ell,p_\ell)$. It is easy to 
see that the canonical commutation relation 
$[\theta_\ell,u_m]=i\hbar\delta_{\ell m}$ holds if 
$[\varphi_\ell,p_m]=i\hbar\delta_{\ell m}$.
The Hamiltonian (3) is greatly simplified if one uses the canonical variables
$(\theta_\ell,u_\ell)$ as \cite{Koyama2}
\begin{equation}
\mathcal{H}=\sum_{\ell=1}^\infty\Bigl\{\frac{E_c}{\hbar^2}\Bigl[
(1+2\alpha)u_\ell^2-2\alpha u_\ell u_{\ell+1}\Bigl]
+E_J\bigl[1-\cos\theta_\ell\bigl]\Bigl\},\label{(6)}
\end{equation}
with $E_c=2\pi de^{\ast 2}/W\epsilon$ and $\alpha=\epsilon\mu^2/sd$. 
Note that the Josephson coupling term is 
diagonal and the coupling between junctions survives only for nearest 
neighbors. The coupling constant $\alpha$ is identical with the one 
introduced in the capacitively-coupled classical intrinsic Josephson 
junctions\cite{Koyama1, Machida1}. In Bi-2212 $\alpha$ takes a value of 
$\sim 0.1-0.2$, depending on the doping level\cite{Preis, Machida1, Machida2}.
Let us discuss the small quantum phase oscillations in the $N$-junction 
system. We consider the case of $N\gg 1$ and neglect the boundary effect 
in the following calculations. For small phase oscillations one can use the 
approximation, $1-\cos\theta_\ell\simeq \frac{1}{2}\theta_\ell^2$. 
Using the periodic boundary condition, we express $u_\ell$ and $\theta_\ell$ 
in terms of the Fourier series expansions as 
$u_\ell=N^{-1/2}\sum_n\hat u_n\exp(i2\pi n\ell/N)$,
$\theta_\ell=N^{-1/2}\sum_n\hat \theta_n\exp(i2\pi n\ell/N)$,
where the Fourier components $(\hat u_n,\hat\theta_n)$ satisfy the 
commutation relation, 
$[\hat\theta_n,\hat u_{-m}]=i\hbar\delta_{\ell m}$. The Hamilotonian (6) 
in the harmonic approximation can be expressed in terms of the Fourier 
components as 
$$
\mathcal{H}=\sum_{n=-n_c}^{n_c}\Bigl\{\frac{E_c}{\hbar^2}\bigl[
1+2\alpha(1-\cos\frac{2\pi n}{N})\bigl]\hat u_n\hat u_{-n}
$$
\begin{equation}
+\frac{1}{2}E_J\hat\theta_n\hat\theta_{-n}\Bigl\},\label{(7)}
\end{equation}
where $n_c$ is given as $N=2n_c+1$. Then, in terms of the creation and 
annihilation operators $c_n$ and $c_n^\dagger$ defined as 
$\hat\theta_n=(c_n+c_{-n}^\dagger)/(\sqrt{2}K_n)$ and $\hat u_n=-i\hbar K_n
(c_n-c_{-n}^\dagger)/\sqrt{2}$ with $K_n=(E_J/\omega_n)^{1/2}$, 
one 
can diagonalize the Hamitonian as 
\begin{equation}
\mathcal{H}=\hbar\sum_{n=-n_c}^{n_c}
\omega_n(c_n^\dagger c_n+\frac{1}{2}),\label{(8)}
\end{equation}
where 
\begin{equation}
\omega_n=\omega_{\rm pl}\sqrt{1+2\alpha(1-\cos\frac{2\pi n}{N})}, \label{(9)}
\end{equation}
with $\omega_{\rm pl}$ being the Josephson-plasma frequency defined as 
$\hbar\omega_{\rm pl}=\sqrt{2E_cE_J}$. 
This phase oscillation mode is identified with the quantum version of the 
longitudinal Josephson plasma. The dispersion relation given in Eq.(9) is 
the same as the classical one\cite{Koyama1, Machida1}. The longitudinal 
Josephson plasma has been observed in the microwave absorption experiments 
in Bi-2212\cite{Matsuda, Kadowaki}. Its frequency $\omega_{\rm pl}$ is 
located in the range of a few hundreds GHz at $T=0$K\cite{Gaifullin}. 
As seen from Eq.(9), we have $N$ plasma modes in the $N$-junction system, 
which is crucially different from the single-junction system, and 
their frequencies range from $\omega_{n=0}=\omega_{\rm pl}$ up to 
$\omega_{n=\pm N/2}=\sqrt{1+4\alpha}\omega_{\rm pl}$. In the case of 
$\alpha\sim 0.1$ (under-doped Bi-2212 case) the relation, 
$\sqrt{1+4\alpha}\omega_{\rm pl}-\omega_{\rm pl}\ll\omega_{\rm pl}$, holds. 
Then, the lowest single-plasmon state is well above the highest zero-plasmon 
state (see Fig.1). From this observation one expects the energy scheme as depicted 
schematically in Fig.1 for the capacitively-coupled Bi-2212 intrinsic 
Josephson junctions in the quantum regime, if the harmonic approximation 
for the phase oscillation modes is assumed to be valid up to the 
single-plasmon state. In this figure we also describe the washboard-like effective 
potential which confines the collective phase oscillation modes. The tilt 
of the potential can be generated by a bias current. Note that in the 
presence of a bias current $I$ the plasma frequency is shifted as 
$\omega_{\rm pl}\rightarrow \omega_{\rm pl}(\tilde I)=\omega_{\rm pl}
(1-\tilde I^2)^{1/4}$ with $\tilde I=I/j_c$ by the effect of the coupling 
term $-E_J\tilde I\sum_\ell\theta_\ell$. 
If $f_{\rm pl}=\omega_{\rm pl}/2\pi\sim 100$GHz, we have the level spacings 
$\hbar\omega_{\rm pl}\sim 4.8$K between zero-plasmon and 
single-plasmon states and $\hbar\omega_{n+1}-\hbar\omega_n\sim 0.23$K 
between adjacent plasma modes for $\alpha=0.1$ and $N=50$. From this 
estimation the quantum effect originating from the discreteness of the 
$n$-plasmon states may be expected at $T\sim 1$K. The level splitting between 
the plasma modes can be seen at temperatures below $\sim 100$mK. 


\begin{figure}
\includegraphics[width=1.0\linewidth]{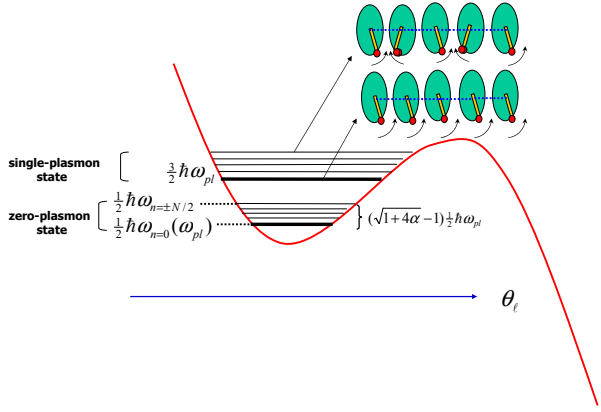}
\caption{\label{fig:epsart} 
Energy-level quantization of the longitudinal collective phase oscillation 
modes in the presence of a bias current. In the $N$-junction system 
single-plasmon states are composed of $N$ oscillation modes which are 
nearly degenerate in the case of $\alpha\ll 1$. The zero-plasmon state is formed by the zero-point motion of the N
plasma modes. }
\end{figure}


Let us now study the MQT in the intrinsic Josephson junctions in which 
the longitudinal Josephson plasma is excited. Suppose that 
the single-plasmon states show the MQT to the voltage state. Since the 
longitudinal plasma oscillations are the coherent motion of all the 
junctions, it is reasonable to assume that the voltage state switched from 
the single-plasmon state is uniform, that is, all the junctions are in the 
voltage state. This switching phenomenon should be discriminated from the 
MQT to the first resistive branch\cite{Inomata}. The quantum switching to 
the first resistive branch is induced by the MQT of the nonlinear localized mode 
such as the discrete breather\cite{Flach} to the localized rotating 
mode\cite{Takeno, Machida1}. The MQT of the discrete breather will be 
discussed in a forthcoming paper. 

In this paper we study the photon-assisted collective MQT in the intrinsic Josephson 
junctions, because the MQT to the uniform resistive state has been observed 
only in the systems under the microwave irradiation up to now\cite{Jin}. 
In the following we focus on the origin of the intriguing $N^2$-dependence 
of the MQT rate recently observed in Bi-2212. For this we consider the 
following third-order anharmonic interaction term that appears in the 
presence of a bias current, 
$$
V_3=-\frac{\gamma}{3!}E_J\tilde I\sum_\ell\theta_\ell^3
$$
\begin{equation}
=-\frac{\gamma E_J\tilde I}{3!\sqrt{N}}\sum_{n_1+n_2+n_3=0,\pm N}
\hat\theta_{n_1}\hat\theta_{n_2}\hat\theta_{n_3},\label{(10)}
\end{equation}
with $\gamma$ being some constant. 
Note that this interaction term induces the coupling among the plasma 
modes, i.e., the mode-mode coupling. The conservation law in Eq.(10), 
$n_1+n_2+n_3=0, \pm N$, comes from the periodic boundary condition. 
In the real systems composed of $N\sim10-100$ junctions this conservation 
law is understood to hold approximately. One expects that this anharmonic 
interaction term becomes important for the excited states, i.e., the 
single-plasmon states, as the bias current is increased. 


\begin{figure}
\includegraphics[width=1.0\linewidth]{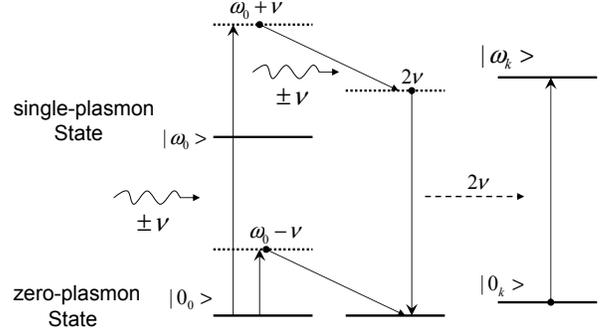}
\caption{
Two-photon process that causes the transition $|0>\rightarrow |\omega_k>$ 
in the third-order perturbation. In this process the virtual plasmon of 
mode $k=0$ with energy $2\nu$, which is excited in the two-photon 
process, excites a plasmon of mode $k$ by the interaction $V_3^k$ when 
$2\nu=\omega_k$. In this figure $|0_0>$ and $|0_k>$ denote, respectively, the vacuum states
of 0th and $k$th plasma oscillations, and $|\omega_0>$ and $|\omega_k>$ the 
single-plasmon states of these oscillations.} 
\end{figure}


Let us investigate the excitations of the plasma mode by microwave 
irradiation, that is, the transition from the ground state $|0>$ to 
$k$th single-plasmon state $|\omega_k>$ (see Fig.2). 
We assume that the coupling between microwave and the phase 
differences is generated by the oscillating current induced by the 
microwave. Then, the interaction with the microwave of frequency $\nu$ 
is described by the term as 
\begin{equation}
V_M(t)=E_J\tilde I_M\sin\nu t 
\sum_{\ell} \theta_\ell
=\sqrt{N}g_M(c_0+c_0^\dagger)\sin\nu t,\label{(11)}
\end{equation}
where $g_M=E_J\tilde I_M/(2\sqrt{2}K_0)$, $\tilde I_M$ being the 
intensity of the normalized oscillating current.
From Eq.(11) one understands that the microwave can excite only the 
uniform plasma mode of $n=0$, $|\omega_0>$, in the one-photon process. 
The transition rate $1/\tau^{(1)}$ in this process is easily derived as 
\begin{equation}
\frac{1}{\tau^{(1)}}=N\pi g_M^2\delta(\omega_{\rm pl}(\tilde I)-\nu).
\label{(12)}
\end{equation}
On the other hand, in the multi-photon processes in which the mode-mode 
coupling is incorporated, the plasma modes with $k\not=0$ can be excited 
as shown in the following. Note that Eq.(10) includes the interaction term, 
\begin{equation}
V_3^k=-\frac{f_k}{\sqrt{N}}c_k^\dagger c_{m_k}c_{m_k}^\dagger,\label{(13)}
\end{equation}
where $f_k=\gamma E_J\tilde I(8K_kK_{m_k}^2)^{-1/2}$ and $m_k$ is the index 
satisfying the relation $k+2m_k=0,\pm N$. Let us study the third-order 
perturbative processes caused by the interaction 
$\mathcal{H}_{\rm int}=V_M(t)+V_3^k+V_3^{k\dagger}$. The two-photon process 
is included in the third-order perturbation and the transition, 
$|0>\rightarrow|\omega_k>$, is possible in the presence of $V_3^k$ as seen 
in Fig.2. Note that $V_M(T)\propto \sqrt{N}$ 
and $V_3^k\propto (\sqrt{N})^{-1}$. Hence, one understands that the 
transition matrix element $<f|i>=<\omega_k|U^{(3)}(t,-\infty)|0>$ in the 
two-photon process is proportional to $\sqrt{N}$, where $U^{(3)}(t,-\infty)$ 
is the third-order time-evolution operator given by
\begin{eqnarray}
U^{(3)}(t,-\infty)=(-\frac{i}{\hbar})^3\int_{-\infty}^t{\rm d}t_1
\int_{-\infty}^{t_1}{\rm d}t_2\int_{-\infty}^{t_2}{\rm d}t_3 \nonumber \\  
{\rm 
e}^{\epsilon(t_1+t_2+t_3)}\mathcal{H}_{\rm int}(t_1) \mathcal{H}_{\rm
int}(t_2) \mathcal{H}_{\rm int}(t_3)\Bigl |_{\epsilon \rightarrow 0},
\end{eqnarray}
and thus the transition rate, 
    $\tau_k^{-1}=\frac{{\rm d}}{{\rm d}t}
             |<\omega_k|U^{(3)}(t,-\infty)|0>|^2$, 
is proportional to $N$. In fact, from the simple calculations we obtain 
\begin{equation}
\frac{1}{\tau_k}=N\frac{\pi(g_M^2f_k)^2}{(\omega_{\rm pl}(\tilde I)-\nu)
(\omega_k+\omega_{\rm pl}(\tilde I)-\nu)}
\delta(\omega_k-2\nu).\label{(14)}
\end{equation}
Then, if the MQT rate of the single-plasmon state $|\omega_k>$ to 
the voltage state is given by $\Gamma_k$, the photon-assisted switching rate 
is obtained as $\tau_k^{-1}\Gamma_k$. In the MQT at temperatures around 
$\sim 1$K the splitting between the plasma modes is smeared out for the 
systems with $N\sim 50$. Then, we find the total switching rate in the 
two-photon process for $\nu\sim \omega_{pl}/2$ as 
\begin{equation}
\Gamma^{(2)}=\sum_{k=-n_c}^{n_c}\frac{1}{\tau_k}\Gamma_k \propto N^2
\frac{\pi(g_M^2f_0)^2}{\omega_{\rm pl}(\tilde I)^2}\Gamma_0, 
\label{(15)}
\end{equation}
which is proportional to $N^2$. The $N^2$-dependence of the switching rate 
is also obtained in the three-photon process. Note that the transition, 
$|0>\rightarrow|\omega_k>$, in the three-photon process is possible in the 
fifth-order perturbative process with respect to $\mathcal{H}_{\rm int}$. 
Then, from Eqs.(10) and (11) we expect the relation, 
$<f|i>\propto (\sqrt{N})^3\times(\sqrt{N})^{-2}=\sqrt{N}$, for the 
three-photon process in the fifth-order perturbation, which is the same as 
in the two-photon process. This observation indicates that the total 
switching rate is also proportional to $N^2$ in the three-photon process. 
Thus, one may conclude that the MQT rate to the uniform resistive state is 
proportional to $N^2$, if the MQT happens collectively in the 
multi-photon process. This result is consistent with the recent experimental 
result in Bi-2212 intrinsic Josephosn junctions\cite{Jin}. Note that the 
$N^2$-dependence dose not appear in the switching rate by way of the 
single-photon process. This is because the single-plasmon state of $k\not=0$ 
cannot be available in the single-photon process. Thus, our theory predicts 
that the switching rate in the single-photon process is proportional to $N$. 
Furthermore, from the result given in Eqs.(15) and (16) it is also predicted 
that the frequency spectrum of the switching rate, i.e., $\Gamma^{(2)}(\nu)$, 
splits into $N/2$ peaks having centers at $\nu=\omega_k/n$ at low 
temperatures, say $T\sim 10$mK, in the $n$-photon process if the damping 
originating from the coupling with the quasiparticles or environment is weak 
enough. 

In summary we have formulated MQT for the collective longitudinal 
Josephson plasma modes in intrinsic Josephson junctions, assuming that 
the switching events to the uniform voltage state occur from the 
single-plasmon states. The scheme of energy-level quantization for the 
collective oscillation modes has been clarified in the case of Bi-2212 
IJJs. In the $N$-junction system the single-plasmon states are composed 
of $N$ different oscillation modes whose frequencies range from 
$\omega_{\rm pl}$ to $\sqrt{1+4\alpha}\omega_{\rm pl}$. These modes can 
be excited by the microwave irradiation in the multi-photon process. Then, 
there are $N$ channels in the MQT of the collective phase oscillation 
modes in the multi-photon process. We have also shown that the transition 
rate to each single-plasmon state in terms of microwave irradiation is 
proportional to $N$. Therefore, the photon-assisted switching rate to the 
uniform voltage state is proportional to $N^2$ in the multi-photon process around 1K. 

The authors thank S.Sato, K.Nakajima, K.Inomata, T.Hatano, H.Kitano, and A.Maeda for discussions 
on experimental results, and A. Tanaka, S. Kawabata, M.Kato, M.Hayashi, H.Ebisawa, and T.Ishida for helpful discussions.

\end{document}